\documentstyle[12pt]{article}
\begin{document}
\newcommand{\be}{\begin{equation}}
\newcommand{\ee}{\end{equation}}
\newcommand{\ba}{\begin{eqnarray}}
\newcommand{\ea}{\end{eqnarray}}
\begin{center}
{\Large Straight-line string with curvature} \\

\vspace{0.2cm}

{\large L.D.Soloviev} \\
{\it Institute for High Energy Physics, 142284, Protvino, Moscow region,
Russia}
\end{center}

\vspace{0.5cm}
\noindent
{\bf Abstract}\\

{\small Classical and quantum solutions for the relativistic
straight-line string with arbitrary dependence on the world surface curvature
are obtained. They differ from the case of the usual Nambu-Goto interaction by
the behaviour of the Regge trajectory which in general can be nonlinear.
Regularization of the action is considered and comparison with
relativistic point with curvature is made.}

\vspace{1cm}

Straight-line string (defined precisely in what follows) is a system with
finite number degrees of freedom. Its quantization does not encounter anomalies
appropriate to the general string. Nevertheless it is an interesting system to
study quantization and various types of interaction. It is a simple extended
relativistic object which can be used to build phenomenological hadron models.
In this paper we solve the problem of straight-line string when its interaction
arbitrarily depends on curvature. We will discuss behaviour of Regge
trajectories in different relativistic models and importance of regularization
for straight-line string interactions.

Let us consider a straight-line string with the action
\be
A=\int\limits^{\tau_2}_{\tau_1}\int\limits^{\sigma_2}_{\sigma_1}F(R/2)\sqrt{-g}
d\sigma d\tau,
\ee
where $\sigma$ and $\tau$ are Poincare invariant space-like and time-like
parameters describing the string and its evolution, $g$ is the determinant of
the string world surface metric, $R$ is the scalar curvature of the surface and
$F(x)$ is an arbitrary function, satisfying a boundary condition. To formulate
this condition we mention that the integrand in (1) should vanish at the ends
of an open string. This follows from variation of (1) with respect to the
ends of the string. We shall assume that the ends of the string are determined
by vanishing of the metric, so that $F$ is less singular than $1/\sqrt{-g}$
when $g\rightarrow 0$. We shall call this case a proper string model to
distinguish it from  a general model which is obtained from a proper string
model by analytic continuation with respect to a parameter in $F$.
For $F(x)=x$ the integrand in (1) is a full derivative.

The straight-line string is described by the vector (for definiteness we
consider 4-dimensional Minkowski space with metric $diag(+1,-1,-1,-1)$)
\be
x^\mu(\sigma,\tau)=r^\mu(\tau)+q^\mu(\tau)f(\sigma,\tau),
\ee
where the Poincare vector $r^\mu$ corresponds to a point on the string, the
translation invariant Lorentz vector $q^\mu$ describes the direction of the
string and the Poincare invariant function $f$ determines the position of
 points on  the string. $r$, $q$ and $f$ are the dynamical variables of the
straight-line string.

It is shown in the Appendix that for the straight-line string (2) the
determinant of the metric and the scalar curvature are given by the simple
expressions:
\be
g=q^2(\dot r_q+\dot q_q f)^2f'^2,
\ee
\be
R/2=-\frac{\dot r^2_q\dot q^2_q-(\dot r_q\dot q_q)^2}{q^2((\dot r_q+\dot q_q f
)^2)^2},
\ee
where dot and prime mean derivatives with respect to $\tau$ and $\sigma$,
respectively, and index $q$ means that the corresponding vector is orthogonal
to $q$:
\be
z_q=z-(zq)q/q^2,\,\,\,z=\dot r,\dot q.
\ee

Putting (3) and (4) into (1) we can derive the Euler-Lagrange equations for $r$
, $q$ and $f$. The equations for $r$ and $q$ contain integrals over $d\sigma$
and equation for $f$ is identically satisfied. Therefore, we prefer to
integrate over $d\sigma$ in (1) at the beginning. The integration reduces to
the integration over $df$ in the limits which correspond to zeros of $g$. As a
result we have the action and the Lagrangian of our model:
\be
A=\int\limits^{\tau_2}_{\tau_1}Ld\tau,
\ee
\be
L=(-\dot n^2)^{1/2}G(l),
\ee
where $n$ is the unit vector in the q-direction:
\be
n=q/(-q^2)^{1/2},
\ee
$l$ is the radius (half of length) of the string at fixed $\tau$
\be
l=(q^2(\dot r^2_q \dot q^2_q-(\dot r_q \dot q_q)^2)/(\dot q^2_q)^2)^{1/2}
\ee
and the function $G$ is obtained from $F$ in (1) by the integration
\be
G(l)=l^2\int\limits^{1}_{0}F((lx)^{-2})(x/(1-x))^{1/2}dx.
\ee
By the assumption this integral exists or is defined through an analytic
 continuation.

To find an extremum of the action (6) and (7) we shall use the Hamilton
method. The momenta canonically conjugate to the "external" and "internal"
coordinates $r$ and $q$ are
\be
p_\mu=-\partial L/\partial \dot r^\mu,\,\,
\pi_\mu=-\partial L/\partial\dot q^\mu.
\ee
Their non-zero Poisson brackets are
\be
\{r^\mu,p_\nu\}=\{q^\mu,\pi_\nu\}=-\delta^\mu_\nu.
\ee
Because of the transformation properties of the action, $r$ and $q$ with
 respect to the Poincare group, the total conserved (independent of $\tau$)
energy-momentum vector of the system coincides with $p$ and the conserved
tensor of the total angular momentum is given by
\be
M^{\mu\nu}=r^\mu p^\nu -r^\nu p^\mu+q^\mu \pi^\nu-q^\nu \pi^\mu.
\ee
The conserved spin tensor is
\be
S^{\mu\nu}=M^{\mu\nu}-(M^{\mu\rho} p^\nu+p^\mu M^{\rho\nu})p_\rho /p^2.
\ee
For the mass and spin of our system we shall use the notations
\be
m=\sqrt{ p^2},\,\,\,S=\sqrt{S^{\mu\nu} S_{\mu\nu} /2}.
\ee

The action (6)-(9) is invariant under three sets of $\tau$-dependent
transformations: shift of $r$ in the direction of $q$, change of $q^2$ and
reparametrization of $\tau$. Therefore there should be three constraints on
the canonical variables. Calculating momenta (11) from (7)-(9) we get
\be
pq=0,\,\,\, \pi q=0,
\ee
\be
S=K(m).
\ee
Here, with constraints (16) fulfilled,
\be
S^2=q^2(\pi ^2-(\pi p)^2/p^2)
\ee
and the function $K$ is determined by the function $G$ by means of the
parametric equations
\be
m=|G'(l)|,
\ee
\be
S=|G(l)-lG'(l)|,
\ee
where prime stands for the derivative with respect to $l$.

Introducing the constraint functions (constraints)
\be
\varphi_1=pq,\,\,\, \varphi_2=\pi q,
\ee
\be
\varphi_3=S-K(m),
\ee
we see that they are constraints of the first kind,i.e. their Poisson brackets
vanish when the constraints are equal to zero. The canonical Hamiltonian of our
system $H_{can}=-p\dot r-\pi \dot q-L=0$, therefore the dynamics of our system
is determined by the Hamiltonian
\be
H=\sum^{3}_{i=1}v_i(\tau)\varphi_i
\ee
and the equation
\be
\dot f=\partial f/\partial t+\{f,H\},
\ee
where $v_i$ is arbitrary and $f$ is a function of the canonical variables.

For the canonical variables themselves equations (24) can be easily solved:
\be
r=2KK'pt_3/m-qt_1,
\ee
\be
q=e^{-t_2}(q_0 \cos T +\pi_p(q^2_0/\pi^2_p)^{1/2}\sin T),
\ee
\be
\pi=e^{t_2}(\pi_p \cos T-q_0(\pi^2_p/q^2_0)^{1/2}\sin T)+pt_1.
\ee
Here $T=2Kt_3+v_0$, $K=K(m)$ is given by (17),(19),(20), $K'=dK/dm$, and $v_0$
is a constant.$q_0$ and $\pi_p$ are constant vectors satisfying (16), $\pi_p$
is orthogonal to $p$, $\pi_p p=0$, and they are connected with spin
\be
S=(q^2_0\pi^2_p)^{1/2}.
\ee
$t_i$, $i=1,2,3$, are arbitrary functions of $\tau$:
\be
t_1=e^{t_2}(c_1+\int v_1 e^{-t_2}d\tau),\,\,\, t_j=c_j+\int v_j d\tau,\,\,\,
j=2,3.
\ee
It is not difficult to show that the first term on the r.h.s. of (25) is the
coordinate of the center of the string. Equating its time component to the
laboratory time
\be
t=2KK'p^0 t_3/m,\,\,\ T=mt/K'p^0 +v_0,
\ee
we see that in the laboratory frame the center of the string is moving with
constant velocity $\vec p/p^0$ and the direction of the string is
rotating in the plane orthogonal to $p^\mu$ and $S^\mu=\epsilon_{\mu\nu\rho
\sigma}p^\nu M^{\rho\sigma}/2m=\epsilon_{\mu\nu\rho\sigma}p^\nu q^\rho_0
\pi^\sigma_p/m$ with the angular velocity
\be
\omega=m/K'p^0=(m/p^0)(dm/dS)=(m/p^0)l^{-1}.
\ee

To quantize our system we use the gauge conditions
\be
p\pi=0,\,\, q^2=\pi^2
\ee
and the tetrad of orthonormal vectors $e_\alpha,\,\,\, \alpha=0,a,\,\,\,
a=1,2,3$
\be
e_0=p/\sqrt{ p^2}, \,\,\, e_\alpha e_\beta=g_{\alpha \beta}.
\ee
Solving the constraint and gauge conditions (16) and (32) we get
\be
q=e_a n^a (\vec S^2)^{1/4},\,\, \pi=e_a [\vec S,\vec n]^a(\vec S^2)^{-1/4},
\ee
\be
S^{\mu \nu}=e^\mu_a e^\nu_b \epsilon_{abc} S^c, \,\,\,  S^2=\vec S^2,
\ee
\be
\vec n^2=1,\,\,  \vec n \vec S = 0.
\ee

To obtain the Poisson (Dirac) brackets of the variables $\vec n, \vec S,r$ and
$p$ one can use method of reduction of the symplectic form [1,2,3]. Using the
initial symplectic form
\be
\omega=dp_\nu \bigwedge dr^\nu+d\pi_\nu \bigwedge dq^\nu
\ee
and equations (34) we get
\be
d\pi_\nu \bigwedge dq^\nu=dp_\nu \bigwedge du^\nu-d[\vec S,\vec n]\bigwedge d
\vec n,
\ee
\be
u^\nu=(1/2)\epsilon_{abc} e_{a\mu}(\partial e^\mu_b /\partial p_\nu)S^c.
\ee
Therefore,
\be
\omega=dp_\nu \bigwedge dz^\nu-d[\vec S,\vec n]\bigwedge d\vec n,
\ee
where
\be
z^\nu=r^\nu+u^\nu.
\ee
Taking into account (36) and calculating the inverse matrix to that in (40)
we get
\be
\{z^\mu,p_\nu\}=-\delta^\mu_\nu,\,\,\, \{S^a,S^b\}=-\epsilon_{abc} S^c,\,\,\,
\{S^a,n^b\}=-\epsilon_{abc} n^c
\ee
Using these Poisson brackets it is easy to see that the Poisson brackets of the
energy-momentum vector $p^\mu$ and the angular momentum tensor $M^{\mu\nu}$
(13),(34) form a representation of the algebra of the Poincare group.

Now it is easy to quantize our system. We replace the variables $z,p,\vec n,
\vec S$ by operators satisfying (36) and the Poisson brackets in (42) by
commutators $\{,\}\rightarrow -i[,]$. The wave function of a physical state
satisfies the equation
\be
\hat\varphi_3\psi\equiv(\sqrt {\hat {\vec S}^2}-K(\sqrt {\hat p^2}))\psi=0.
\ee
In the representation where $\hat p$ and $\hat{\vec n}$ are diagonal a basis of
physical states is given by
\be
\psi_{\vec k SS^3}(p,\vec n)=c\delta (\vec p-\vec k)
\delta (p^0-\sqrt{\vec k^2+m^2_S}) Y^S_{S^3}(\vec n),
\ee
where $Y^S_{S^3}(\vec n)$ is the spin eigenfunction,
\be
\sqrt{S(S+1)}=K(m_S)
\ee
and $S$ is a non-negative integer. It is important to notice that the operator
of the angular momentum obtained from (13),(34),(39),(41) does not contain
noncommuting multipliers. Therefore, the commutators of $\hat M^{\mu\nu}$ and
$\hat p^\rho$ coincide (up to $i$) with their Poisson brackets and form a
representation of the algebra of the Poincare group,i.e. our quantized theory
is Poincare invariant.

Let us discuss the obtained results.First we mention the problem of
regularization of integrals (1) and (10). Even in a trivial case $F(x)=x$, when
Lagrangian (1) is a full derivative, we encounter divergence in integral (10).
The boundary conditions at the ends of the string are not fulfilled either.
We need a regularization, for instance $F(x)=x^\gamma$ to get (52), from where
we get zero action at $\gamma\rightarrow 1$, as it should be. Let us consider
from this point of view the Polyakov interaction[5] in the straight-line
approximation. Polyakov suggested a term in the string action proportional to
\be
\int\int P\sqrt{-g}d\sigma d\tau,\,\,\,   P = \sum^{2}_{a=1}(Spb^{(a)})^2,
\ee
where $b^{(a)}$ is the second quadratic form of the string world surface
\be
b^{(a)}_{ik}=(\partial^2 x/\partial\sigma^i\partial\sigma^k, n^{(a)}),\,\,\,
\sigma^1=\sigma,\,\,\,  \sigma^2=\tau,
\ee
$n^{(a)}, a=1,2$ being the orthonormal vectors orthogonal to the surface and
$Sp b^{(a)}=g^{ik}b^{(a)}_{ki}$. In the conformal gauge
\be
P=(\ddot x- x'')^2/g.
\ee
In the straight-line approximation (2)
\be
P=-\frac{(a+bf)^2}{[(\dot r_q+\dot q_q f)^2]^2},\,\,\,
a=\ddot r_q-2\dot q_q(\dot r q)/q^2,\,\,\,  b=a(r\rightarrow q)
\ee
(see Appendix) and the integral over $d\sigma (df)$ in (46) diverges.
Regularizing (49) by replacing its denominator by its $\gamma$th power (it is
not a gauge invariant regularization, but the limit at $\gamma\rightarrow 1$
is gauge invariant) we can integrate over $df$. Going to the limit $\gamma=1$
we get a finite Lagrangian
\be
L=-\pi(q^2(\dot q^2_q)^{-3})^{1/2}b^2,
\ee
which does not depend on the derivatives of $r$.This result is not satisfactory
because it corresponds to zero energy-momentum vector $p=0$. Therefore the
straight-line approximation (or the aforementioned regularization)
is not valid  for this particular interaction.

Next we see that our relativistic model (7)
possesses a Regge trajectory, i.e. spin of the system depends on its mass and
not on other continuous dynamical variables. The origin of this phenomenon is
the high symmetry of the Lagrangian (7): it is invariant not only under
reparametrization, but also under a shift of $r$ in the $q$-direction and under
renormalization of $q$. We can compare our model with the relativistic particle
with curvature [4] $L=f(k)(\dot x^2)^{1/2},\,\,\, k=(-\ddot x^2)^{1/2}/
\dot x^2$.  Here $\dot x$ can be treated as a new coordinate, so this system
has the same number of degrees of freedom as (7). But because of absence of the
aforementioned symmetry and corresponding constraints spin of this system
depends not only on its mass, but also on two canonically conjugate variables
which vary from $-\infty$ to $+\infty$. Only for $f(k)=k$ there exists an
additional constraint and spin depends on mass only, decreasing with the
increase of mass.

In our model, depending on the function $F$ in (1), the behaviour of Regge
trajectory may be nonlinear. It is well known that the linear with respect to
$m^2$ rise of Regge trajectories of hadrons composed with the same quarks is a
remarkable experimental fact lacking complete theoretical understanding. The
straight-line string with the Nambu-Goto interaction $F(x)=const$ reproduces
this behaviour, while many other relativistic models give decreasing
trajectories. We see now that in a general relativistic model (7) the behaviour
of the Regge trajectory may be nonlinear depending on the function $G$ or $F$.
To see this and the limitations following from the string origin of (7) let us
consider an example
\be
F(x)=-cx^\gamma.
\ee
For the proper string model satisfying boundary conditions
 it is necessary that $\gamma\leq 1/4$. On the other hand, the integrals
in (1) and (10) exist if $\gamma<3/4$ and can be defined by the analytic
continuation in $\gamma$ for $\gamma\not= (2n+3)/4,\,\,\, n=0,1,2...$. The
function  $G(l)$ in (10) is now
\be
G(l)=-c\pi^{1/2}\frac{\Gamma(3/2-2\gamma)}{\Gamma(2-2\gamma)}l^{2-2\gamma}=
-bl^{2-2\gamma}.
\ee
For the total energy of the system to be positive it is necessary that
$b(1-\gamma)>0$. The Regge trajectory (45) is given by
\be
\sqrt{S(S+1)}=\beta(m^2)^a,
\ee
\be
a=\frac{1-\gamma}{1-2\gamma},
\ee
\be
\beta=|b(1-2\gamma)|(4b(1-\gamma))^{-2a}.
\ee
We see that for the proper string model $1/2<a\leq 3/2$. For a general
relativistic model of type (51) $(\gamma>1/4)$ the Regge trajectory can
increase or decrease with $a\not=0$ and $a\not=  (2n-1)/(4n+2),\,\,\,
n=0,1,2,...$.For $F(x)=c+bx^{-A}$, where $A>0$, spin $S$ (or $\sqrt{S(S+1)}$
in quantum case) is proportional to $m^2$ for small mass and to $m^a,
a=(A+1)/(A+1/2)$ for large mass.

It is easy to see that for $m\rightarrow\infty$ spin is proportional to
$m^a,\,\,\, a\geq 1$ for any choice of $G(l)$. We can not have $S\rightarrow
const$ in this limit in our model which is a model of " permanent confinement".

We see that the linear rise of Regge trajectories corresponds to a specific
quark-gluon dynamics. On general grounds one can expect deviations from this
behaviour, therefore further experimental study of Regge trajectories at
higher masses and spins is of considerable theoretical interest. The model
considered can serve for a phenomenological description of experimental
trajectories.

{}.

\section{Appendix. Calculation of $g$, $R$ and $A$.}
Calculation of $g=det\parallel(\partial_i x\partial_j x)\parallel,
\partial_i=\partial /\partial \sigma^i$ for the straight-line ansatz (2) is
straightforward with the result (3) in the text. We can use of course the
comformal gauge $\dot x x'=0, \rho=\dot x^2=-x'^2$ in which
$\dot f=-(\dot r q+\dot q qf)/q^2$ and $-q^2f'^2=(\dot r_q+\dot q_q f)^2$, so
that $g=-\rho^2=-(q^2f'^2)^2=q^2 f'^2(\dot r_q+\dot q_q f)^2$,
because we need $f'$ to integrate over $df$.
The use of the comformal gauge is very helpful for calculating $R$. In this
gauge
\be
R=(1/\rho)[(\dot\rho/\rho)^. -(\rho'/\rho)'].
\ee
Differentiating $\dot f$ with respect to $\sigma$ we get
$\dot f'=-\dot q qf'/q^2$, so that $\dot\rho=(-q^2f'^2)^. =0$. Now
$f''=-(\dot r_q\dot q_q+\dot q^2_q f)/q^2, f'''=-\dot q^2_q f'/q^2$, so that
$R=[(f'^2)'/f'^2]'/q^2f'^2=2(f'''f'-f''^2)/q^2f'^4$  and we get the equation
(4) in the text.

The same is applied to $A$ in (52). All the derivatives of $f$ can be expressed
through $f$ and we get (53) in the text.

\end{document}